\documentclass[12pt]{iopart}
\usepackage{iopams}
\input epsf.sty

\begin{document}

\input epsf.sty

\def\g{\gamma}
\def\r{\rho}
\def\av#1{\langle#1\rangle}
\def\pf{P_{\rm F}}
\def\pr{P_{\rm R}}
\def\F#1{{\cal F}\left[#1\right]}

\title{Aging processes in reversible reaction-diffusion systems: Monte Carlo simulations}

\author{Nasrin Afzal$^1$, Justin Waugh$^2$, and Michel Pleimling$^1$}


\address{$^1$Department of Physics, Virginia Tech, Blacksburg, Virginia
24061-0435, USA}
\address{$^2$Department of Physics, University of Colorado at Boulder, Boulder, Colorado 80309-0390, USA}


\begin{abstract}
Reaction-diffusion systems with reversible reactions generically display power-law relaxation towards 
chemical equilibrium. In this work we investigate through numerical simulations aging processes
that characterize the non-equilibrium relaxation. Studying a model which excludes multiple occupancy of 
a site, we find that the scaling behavior of the two-time correlation and response functions are
similar to that discovered previously in an exactly solvable version with no restrictions on the
occupation numbers. Especially, we find that the scaling of the response depends on whether the
perturbation conserves a certain quantity or not. Our results point to a high degree of universality
in relaxation processes taking place in diffusion-limited systems with reversible reactions.
\end{abstract}
\maketitle

\section{Introduction}
Slow dynamics and aging processes are encountered in a huge variety of systems. The best known examples can of course be
found in glassy systems, i.e. structural glasses and spin glasses, but intensive research in the last decade has revealed
that this type of processes are ubiquitous and can be found in magnetic systems, in non-equilibrium growth or in
diffusion-limited reaction systems. A recent overview of the field, with a focus on aging phenomena in non-glassy systems,
can be found in \cite{henple}. Slow dynamics in non-glassy systems is often much easier to understand than in the more
complex glassy systems, and much progress in our understanding of the general properties of
aging processes has been achieved through the study of rather simple model systems with slow dynamics.

Reaction-diffusion systems have greatly contributed to our current understanding of the physics far from equilibrium.
This is especially true for non-equilibrium phase transitions as encountered in systems with irreversible reactions and 
absorbing states \cite{book1,Odo08}. In this context, the study of relaxation and aging phenomena remains in its early stages. Still,
some interesting results have been found recently. As is well known, power-law relaxation, the hallmark of slow dynamics,
is observed in diffusion-limited irreversible reactions at their (non-equilibrium) phase transition points. Consequently,
the first studies of aging in reaction-diffusion systems were restricted to these non-equilibrium critical points that are
characterized by the absence of detailed balance \cite{Ens04,Ram04,Bau05,Hin06,Odo06,Bau06,Bau07,Tak07,Hen07,Dur09,Dur10,Dur11}. 
The analysis of two-time correlators $C(t,s)$ and responses $R(t,s)$ at an absorbing phase transition revealed a phenomenology similar
to that observed at an equilibrium critical point. Both quantities exhibit standard scaling forms, 
\begin{equation} \label{eq:scal1}
C(t,s) = s^{-b} f_c(t/s) , ~~~ R(t,s) = s^{-1-a} f_R(t/s)
\end{equation}
where the scaling functions display a simple power-law behavior for large arguments, $y = t/s \gg 1$:
\begin{equation} \label{eq:scal2}
f_C(y) \sim y^{- \lambda_C/z}, ~~~ f_R(y) \sim y^{- \lambda_R/z}~.
\end{equation}
Here $s$ and $t > s$ are two different times called waiting and observation times, and $z$
is the dynamical exponent. The exponents $a$, $b$, $\lambda_C$, and $\lambda_R$ are non-equilibrium exponents that
govern the scaling properties of the two-time quantities. However, these studies also revealed some remarkable differences to the
aging properties of systems relaxing towards an equilibrium steady state. Most notably, the exponents $a$ and $b$,
which for a system relaxing towards equilibrium are identical and related to a static critical exponent \cite{henple},
can be different when detailed balance is broken. In some cases this difference can be understood through symmetry
properties of the model, as for example for the contact process \cite{Bau07}, in other cases no simple explication
seems to exist.

In order to observe slow dynamics at an absorbing phase transition, a fine-tuning of the system parameters is
obligatory. This is, however, different for systems with reversible reactions, as these systems generically display power-law
relaxation, independent of the values of the reaction and diffusion rates \cite{Zel77a,Zel77b,Kan85,Osh89a,Osh89b,
Agm94,Agm95,Rey99,Gop00,Agm00,Gop02}. Studies of excited-state
proton transfer reactions provided experimental verifications of this theoretically predicted behavior \cite{Hup92,Sol01,Pin01}. 
This of course makes the diffusion-limited systems with reversible reactions very attractive for a study of
aging processes.

In \cite{Elg08} a first step was undertaken in that direction where the non-equilibrium dynamical properties of
some exactly solvable models were studied. This study indeed revealed the presence of the standard scaling forms
(\ref{eq:scal1}) and (\ref{eq:scal2}) for correlation and response functions, and this for any values of the system
parameters. Surprisingly, however, the scaling of the response function was found to strongly depend on whether
or not some specific quantity was kept constant when perturbing the system. 
More precisely, the study of the exactly solvable models showed that the  breaking of conserved quantities yields
an additional contribution to the response function that is waiting time independent.
This is an interesting result, as it
highlights the importance of conserved quantities during non-equilibrium relaxation.

The models studied in \cite{Elg08} are to some extend specialized in order to be exactly solvable. It is therefore
of interest to understand which of the properties found in \cite{Elg08} are generic to systems with reversible reactions
and which depend on special features of the studied models. Thus, in order to be soluble, it was assumed that
the possible number of particles at every lattice site was unrestricted. In addition, as initial condition
an uncorrelated Poisson distribution on each lattice site and for each particle species was assumed.
In this paper we study a system composed of two types of particles that undergo reversible reactions while 
diffusing on a lattice. We thereby only allow single site occupation and prepare the system by randomly distributing
on the lattice a given number of particles of every type. During the relaxation process we study the two-time 
autocorrelation function as well as different two-time response functions.

The paper is organized in the following way. In the next section we discuss in more detail our model and introduce
the various quantities that are monitored during the relaxation process. Section 3 is devoted to our numerical
results. Focusing on the autocorrelation and autoresponses, we study their scaling properties, and this both
in one and two dimensions. Finally, we give our conclusions in section 4.

\section{Model and quantities}
We consider one- and two-dimensional lattices on which particles of different types, called $A$ and $C$, diffuse
and interact. The data discussed in the following have mainly been obtained for systems of
linear extend of $L =10000$ in $d=1$ and $L =100$ in $d=2$. We carefully checked that
no finite-size effects show up in our quantities for these sizes. The diffusion of particles is realized by jumping
to unoccupied nearest neighbor sites.
The reaction scheme considered here is given by
\begin{equation}
A+A \rightleftarrows C
\end{equation}
where two neighboring $A$ particles coalesce to form a $C$ particle, whereas a $C$ particle
can decompose into two $A$ particles, provided that one of the neighboring sites is empty. In the
simulations, we realize this scheme in the following way. We first select randomly a site and a direction. 
If the selected site is occupied by a $C$ particle and 
if the neighboring site in the selected direction is empty, then the $C$ particle is hopping in that direction
with probability $D_c$ or it is replaced by two neighboring $A$ particles with probability $\mu$. If, on the other
hand, an $A$ particle sits on the selected site, the possible action will depend on whether the neighboring site
in the selected direction is either empty or occupied by an $A$ particle. In the first case we move the particle
with probability $D_a$, in the second case the two $A$ particles are replaced with probability $\lambda$ by a $C$
particle that we put on the initially selected site. For all other cases, no action takes place.

We prepare our system at time $t=0$ ''microcanonically'' by distributing $N_A(0)$ $A$ particles and $N_C(0)$ $C$ particles
randomly on our lattice. Preparing the system in this strict way is in fact very important as the quantity
\begin{equation} \label{eq:K}
K = N_A(0) + 2 N_C (0) = N_A(t) + 2 N_C(t)
\end{equation} 
is independent of time, making it a conserved quantity \cite{Rey99}. Here $N_A(t)$ and $N_C(t)$ are the numbers of $A$ and
$C$ particles at time $t$. It follows from the presence of this conserved quantity that the steady-state
particle densities $\rho_{A,S} = N_A(t \longrightarrow \infty)/N$ and $\rho_{C,S} = N_C(t \longrightarrow
\infty)/N$, with $N = L$ in one dimension and $N = L \times L$ in two dimensions, depend on the initial 
preparation of the system, besides depending on the values of the model parameters.

In our simulations we monitor both the particle density as well as the two-time correlation function. Defining the
time dependent occupation number $n_C^i(t)$ of site $i$ as having the value 1 when that site is occupied at time
$t$ by a $C$ particle and zero otherwise, the particle density of $C$ particles is given by
\begin{equation}
\rho_C(t) = \sum\limits_i n_C^i(t)/N~.
\end{equation}
Due to the conserved quantity $K$, the particle density of $A$ particles is obtained directly through the
relation
\begin{equation}
\rho_A(t)= K/N - 2 \rho_C(t)~.
\end{equation}
Insights into the scaling properties during the relaxation process can be gained by studying the two-time
connected autocorrelation function for $C$ particles,
\begin{equation}
C_C(t,s) = \langle \frac{1}{N} \sum\limits_i n_C^i(t) n_C^i(s) \rangle - \langle \rho_C(t)\rangle \langle \rho_C(s) \rangle~
\end{equation}
where the notation $\langle \cdots  \rangle$ indicates averages over both initial conditions
and realizations of the noise.
In a similar way we define the two-time connected autocorrelation function for $A$ particles, $C_A(t,s)$.

For the two-time response function different cases can be distinguished. Firstly, one can perturb the system
in such a way that the quantity $K$ does not change its value. This can be achieved, for example, by 
selecting an empty site and adding with probability $r_K$ a new $C$ particle to the system
while removing at the same time pairs of $A$ particles.
Obviously, we do not think that this is a realistic scenario that
can be easily achieved in an experiment. Still, this scenario can be studied theoretically and might yield
important insights into the role played by conserved quantities during relaxation processes. Secondly, one
can also perturb the system such that $K$ is no longer conserved. An obvious way of doing that is to inject
on randomly selected empty sites
additional particles of only one type (say, $C$ particles) with probability $r$. This will change the value of $K$, and the
system will relax to a steady state whose particle densities differ from those of the unperturbed system.

In our simulations we have implemented the calculation of the responses in the following way. Initially,
we prepare the system in a random state with fixed numbers of $A$ and $C$ particles, $N_A(0)$ and $N_C(0)$.
This fixes the value of $K$ at time $t=0$. Calling
$M_K(t,s)$ the response of the system to a perturbation that keeps $K$ constant, we compute the difference
between the average densities with and without perturbation:
\begin{equation} \label{res_K}
M_K(t,s) = \left[ \langle \rho_C^p(t,s) \rangle - \langle \rho_C(t) \rangle \right] /r_K~.
\end{equation}
Here $\rho_C^p(t,s)$ is the density of $C$ particles at time $t$ when the perturbation is removed
at time $s < t$. Alternatively, we could define the response through the change in density of
$A$ particles. We carefully checked that our conclusions do not depend on our choice of the particle density.
For that reason we will restrict ourselves in the next section to a discussion of the change in 
the density of $C$ particles.

The situation is remarkably different when considering the response to the injection of particles of a single
type. In that case $K$ is no longer conserved, and the system, after removal of the perturbation,
relaxes to a different steady state than the unperturbed system. In fact, depending on the realization of the
noise, i.e. on the sequence of random numbers, different runs end up with different values of $K$, yielding 
different asymptotic steady states. Therefore, in order to be able to monitor the relaxation of every perturbed run to its
asymptotic steady state, we need to compute for every run the difference in particle densities between the perturbed
and an unperturbed run that both yield the same (unknown) steady state. In order to do that we proceed as follows.
For every run we prepare the system as before and then inject additional particles until time $s$. After removing the
perturbation we determine the particle densities $\rho_C^p(s,s)$ and $\rho_A^p(s,s)$ as well as
the new value of $K$ which will remain constant for the remainder of the
simulation. Having determined $K$, we then do a second, unperturbed run, where we start with a disordered
initial state with particle densities 
\footnote{Using different initial particle densities but the same value of $K$ yields the same response
in the scaling limit.}
$\rho_C(0)=\rho_C^p(s,s)$ and $\rho_A(0) = \rho_A^p(s,s)$, i.e. this
run takes place with the same value of $K$ as realized in the perturbed run once the perturbation has been
removed.
The response to the injection of $C$ particles is then calculated as the averaged difference 
between the densities with and without perturbation:
\begin{equation} \label{res_C}
M_C(t,s) = \langle \rho_C^p(t,s)  -  \rho^K_C(t) \rangle/r~.
\end{equation}
Here $\rho^K_C(t)$
is the time-dependent particle density of the unperturbed run at the value of $K$ that is obtained from 
the corresponding perturbed run.

In this study we are only interested in the linear response regime. We carefully checked that for the
values of the probabilities $r$ and $r_K$ considered in the next section the linear response regime
prevails.

\section{Simulation results}
In the following we discuss the relaxation and aging processes that take place in our system. In our systematic
study we considered diffusion constants $D_a$ and $D_c$ between 0.05 and 0.5, whereas the reaction probabilities $\lambda$ and
$\mu$ were varied between 0.1 and 0.7. We also considered a vast range of initial conditions in order to study
the dependence of our results on the value of $K$. In addition, for a fixed value of $K$ we typically considered
three sets of initial conditions that differ by the number of $A$ and $C$ particles deposited on the lattice.
Our main results for the time-dependent particle densities as well as for the two-time autocorrelation
and response functions are summarized in the following subsections. These results are
confronted with the analytical calculations obtained for the related model studied in \cite{Rey99,Elg08}.

\subsection{Particle densities}
As discussed in many theoretical studies \cite{Zel77a,Zel77b,Kan85,Osh89a,Osh89b,
Agm94,Agm95,Rey99,Gop00,Agm00,Gop02} and as verified in some experiments \cite{Hup92,Sol01,Pin01}, systems with
reversible chemical reactions are characterized by slow dynamics such that the particle densities approach their
steady-state values with a power-law in time. More specifically, for the exactly solvable models
without site restriction studied in \cite{Rey99,Elg08}
one finds that particle densities asymptotically behave like
\begin{equation}
\rho(t)-\rho_S \sim t^{-d/2}
\end{equation}
for any dimension $d$. Explicit expressions
have been found for the stationary particle densities which reveal a dependence only on the value of the conserved
quantity $K$ and the ratio $\mu/\lambda$ of the reaction probabilities \cite{Rey99}.
In fact, the steady state is a (chemical) equilibrium state, and the expressions for the stationary particle densities
follow from the solution of the detailed balance condition of
the master equation \cite{Rey99}.

\begin{figure}[h]
\centerline{\epsfxsize=4.25in\ \epsfbox{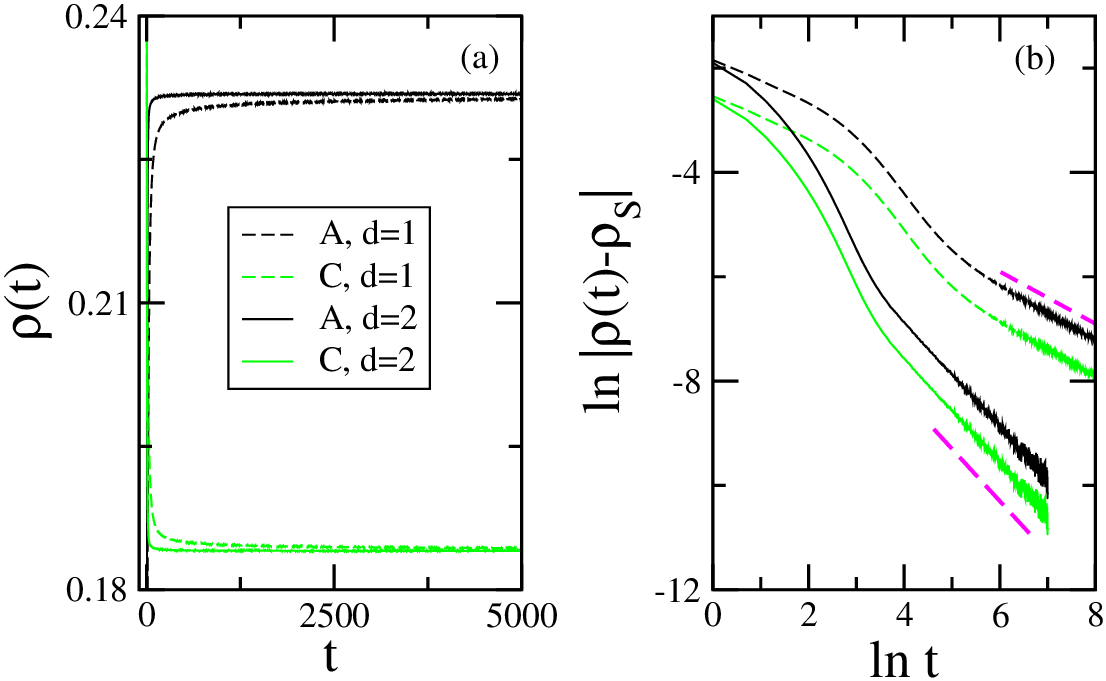}}
\caption{(a) Time evolution of the densities of $A$ and $C$ particles
and (b) approach to stationarity of the particle densities in one and two dimensions.
The linear sizes of the systems are 
$L=10000$ in $d=1$ and $L=100$ in $d=2$, the common probabilities being
$D_a = D_c = 0.1$ and $\mu = \lambda = 0.5$. Initially the systems consist only of $C$ particles that
randomly occupy 30\% of the lattice sites. The approach to stationarity is governed by a power law
with exponent $d/2$, as indicated by the dashed straight segments. The data shown here result from averaging
over $400000$ independent runs.}
\label{fig1}
\end{figure}

In Figure \ref{fig1} we show the typical time evolution of the particle densities for our system. After an initial fast change,
the particle densities rapidly evolve towards a regime where the approach to the steady state is algebraic.
Figure 1 shows two cases with identical reaction and diffusion probabilities as well as with identical initial densities, the
only difference being the dimensionality of the lattice. In both cases, the asymptotic particle densities are found to be identical and
independent of whether a line or a square lattice is considered, see Figure \ref{fig1}a. Changing the diffusion and reaction probabilities in a systematic
way, we find that the stationary particle densities remain
unchanged when the values of the diffusion constants are changed. In fact, and this is in agreement
with the expressions obtained for the exactly solvable case in \cite{Rey99}, the steady-state values of the
particle densities are completely fixed by the values of $K$ and of the ratio $\mu/\lambda$. 
This should not come as a surprise as also in our case 
the steady state is an equilibrium state, due to the reversibility of the reactions.
We also observe, see Figure \ref{fig1}b, that the approach to
stationarity is governed by the exponent $d/2$, as it is the case for the corresponding exactly solvable model \cite{Rey99}.

\subsection{Autocorrelation}

\begin{figure}[h]
\centerline{\epsfxsize=4.25in\ \epsfbox{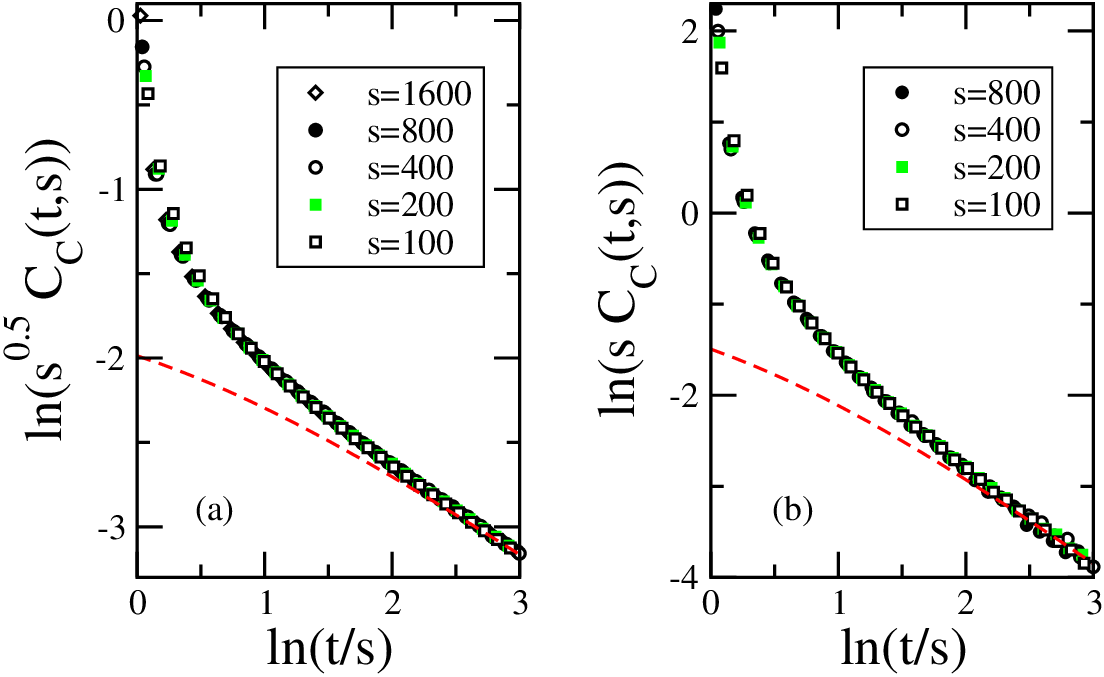}}
\caption{The scaling of the autocorrelation function $C_C(t,s)$ for $C$ particles in (a) $d=1$ and (b) $d=2$. The diffusion constants
are $D_a = D_c = 0.05$, whereas the reaction probabilities are $\mu = \lambda = 0.5$, the initial state being
composed only of $C$ particles that
randomly occupy 30\% of the lattice sites. The red dashed lines show the asymptotic scaling functions obtained for the
exactly solvable model studied in \cite{Elg08}, see equation (\ref{eq_cor_elg}). The data result from averaging over at
least 30000 independent runs.}
\label{fig2}
\end{figure}

The scaling of the two-time autocorrelation function in one and two dimensions is shown in Figure \ref{fig2} 
for the $C$ particles (similar results are obtained when looking at $A$ particles).
In all cases we obtain the standard aging scaling (\ref{eq:scal1}) and (\ref{eq:scal2}) with exponents $b = d/2$ and $\lambda_C/z=d/2$. 
Comparing the scaling functions for different values of the probabilities and different initial densities but fixed
dimensionality, one observes that small differences, present for small values of $t/s$, rapidly vanish when
$t/s$ increases. Disregarding these finite time corrections, one recovers for a fixed value of $d$ a common
scaling function for all probabilities and initial states.

It is tempting to compare our scaling functions with those obtained when site restriction is not imposed. As
shown in \cite{Elg08} the scaling function is then given in leading order by
\begin{equation} \label{eq_cor_elg}
C_C(t,s) = A \, s^{-d/2} \left( t/s + 1 \right)^{-d/2}
\end{equation}
where the amplitude $A$ depends on reaction probabilities as well as on initial conditions and the steady-state particle
densities. One therefore obtains for both models the same values of the exponents $a$ and $\lambda_C/z$. The
expression (\ref{eq_cor_elg}) however only slowly approaches the scaling function obtained in the present study, as
shown in Figure \ref{fig2}. In fact, equation (\ref{eq_cor_elg}) only gives the asymptotic scaling function, valid
in the limit $s \longrightarrow \infty$ and $t\longrightarrow \infty$, with $t> s$. Subleading correction terms, 
which have not been identified systematically for the exactly solvable model
\cite{Elg08}, can not be neglected on the time scale of our simulations.

\subsection{Response functions}

Our main motivation for the present study was the surprising observation in \cite{Elg08} that the scaling function
of the autoresponse strongly depends on how the system is perturbed. As already mentioned in the introduction,
one obtains for the exactly solvable models studied in \cite{Elg08} expressions for the responses that depend on whether
or not certain quantities are conserved during the perturbation.
Our main aim in the following is to verify whether this is a generic behavior or whether this follows from the
special properties of these exactly solvable models.

We first remark that we wish in the following to understand the linear response regime. For that we carefully selected
the probabilities for injecting additional particles by determining the range of values that leads to a linear
dependence of the response on the values of the injection probabilities. We discuss in the following only cases that
are well within that range.

\begin{figure}[h]
\centerline{\epsfxsize=4.25in\ \epsfbox{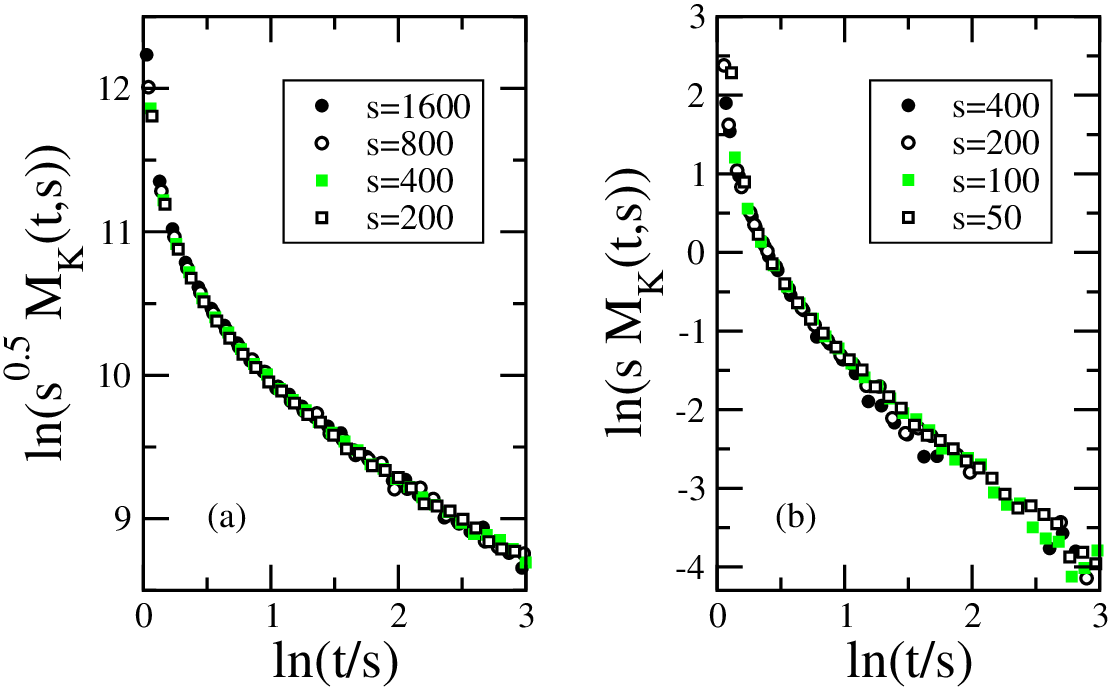}}
\caption{The scaling of the time integrated response $M_K(t,s)$, see equation (\ref{res_K}), in (a) $d=1$ and (b) $d=2$. 
The diffusion constants
are $D_a = D_c = 0.5$, whereas the reaction probabilities are $\mu = \lambda = 0.5$. The probabilities for injecting $C$ particles
and removing pairs of $A$ particles
are $r_K = 0.01$ in $d=1$ and $r_K = 0.05$ in $d=2$. Initially 14\% of the sites are occupied
by $A$ particles, whereas $C$ particles are randomly deposited on 23\% of the sites. The data shown here have been obtained after
averaging over typically one million independent runs.}
\label{fig3}
\end{figure}

We start by showing in Figure \ref{fig3} the time-integrated response (\ref{res_K}) where the injection of a $C$ particle
is accompanied by the removal of two $A$ particles such that $K$ remains constant. We recover a simple aging behavior,
with exponents $a = d/2$ and $\lambda_R/z = d/2$, similar to what is obtained for this response when studying
the model without site restriction. \footnote{Note that in \cite{Elg08}
it is also the time integrated response that is investigated. This is not correctly stated in that paper.}
This scaling behavior of the integrated response is therefore similar to that observed in many systems relaxing
toward their steady state while being characterized by a single time-dependent length scale that increases
algebraically with time \cite{henple}. 

\begin{figure}[h]
\centerline{\epsfxsize=4.25in\ \epsfbox{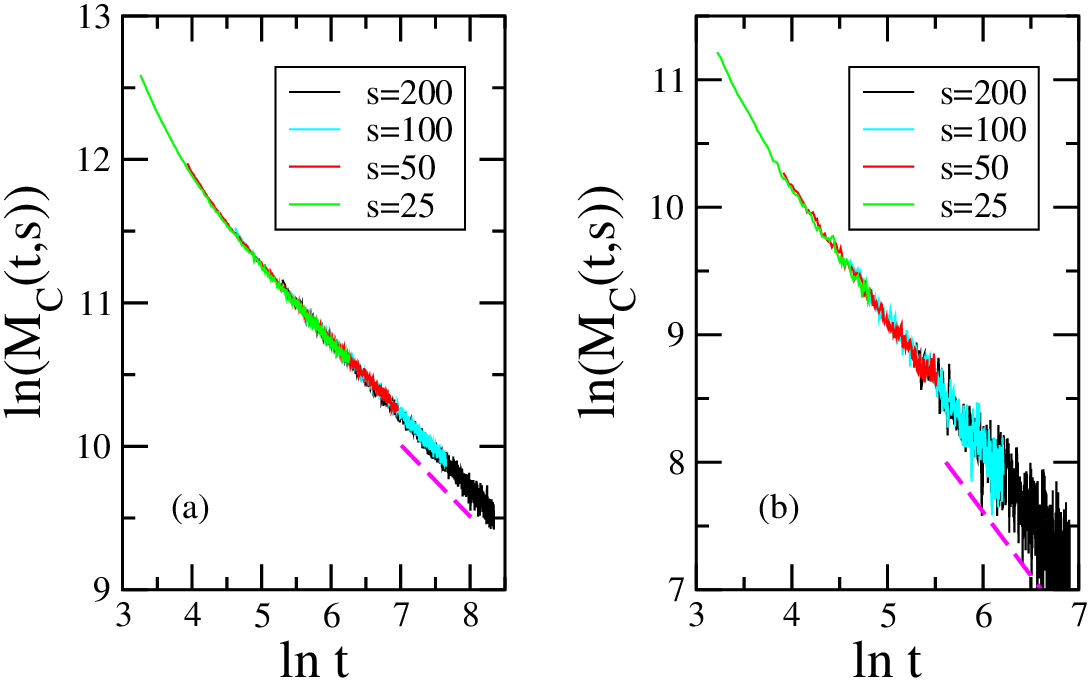}}
\caption{The scaling of the time integrated response $M_C(t,s)$, see equation (\ref{res_C}), in (a) $d=1$ and (b) $d=2$.
The diffusion constants
are $D_a = D_c = 0.05$, whereas the reaction probabilities are $\mu = \lambda = 0.5$. The injection probabilities of $C$ particles
is $r = 0.0001$ in both cases. Initially 14\% of the sites are occupied
by $A$ particles, whereas $C$ particles are randomly deposited on 23\% of the sites. The data shown here have been obtained after
averaging over typically 250000 independent runs. Note that the response is waiting time independent and that
the decay is algebraic for large times, with an exponent compatible with $d/2$, as indicated by the dashed lines.}
\label{fig4}
\end{figure}

The scaling of $M_K(t,s)$ has to be contrasted with the scaling of $M_C(t,s)$ where the perturbation consists in
injecting additional $C$ particles. As shown in Figure \ref{fig4} $M_C(t,s)$ is {\it independent} of the waiting time and, after
some initial faster decay, rapidly exhibits a power-law behavior,
\begin{equation}
M_C(t,s) \sim t^{-\lambda/z}~,
\end{equation}
where the value of $\lambda/z$ is compatible with $d/2$,
as indicated by the dashed lines in the figure. 
For the model studied in \cite{Elg08} the time integrated
response to the injections of only $C$ particles is the sum of two terms, one proportional to $t^{-d/2}$ and one
proportional to $(t-s)^{-d/2}$, see equation (41) in \cite{Elg08}. 
The absence of the second term in our simulations can be understood when recalling the definition of our response.
Having determined the value of $K$ at the end of the perturbation, we start a second, unperturbed, run with that value of
$K$ and after $s$ time steps we start measuring the difference in particle densities between the initial, perturbed, run and
the new, unperturbed, run. Therefore, we always compare densities that relax to the same steady-state value with
the same power-law but with different amplitudes. Because of this way of measuring the response, an independence on the
waiting time can be expected.


Even so the behavior of our response differs from that in the model without site restriction, the main conclusion
drawn in \cite{Elg08} is still valid: the scaling properties of responses change when otherwise conserved quantities 
are changed due to the perturbation. 
More precisely, the appearance of a waiting time independent part in the response is due to the change of $K$
during the perturbation. 
This is a mathematical precise statement for the exactly solvable model
with reversible reactions studied in \cite{Elg08}. Similar calculations are not possible for the model studied here,
but the emergence of a waiting time independent response suggests that the same mechanism is in place.
Our results therefore suggest 
that this is a more general property of aging processes that take place in systems with conserved quantities.

Finally, let us put this result into a broader context by comparing it to results obtained in other systems.
Generally, one expects a response function to depend on the way the system has been perturbed (see, for example,
\cite{Cor10}). However, in all
cases known to us the scaling exponents are found to be independent of the perturbation. This is different for
the reversible reaction-diffusion models, where the breaking of the conserved quantity yields a contribution
of the response with a different dependence on the waiting time.

\section{Conclusion}

We have discussed in this paper the non-equilibrium relaxation in a reaction-diffusion model characterized by reversible
reactions. In contrast to systems with irreversible reactions, systems with reversible reactions generically display
slow dynamics and simple aging, independent of the values of the reaction and diffusion probabilities. Comparing our results with those
obtained previously for a model with site restriction, we note that the approach to stationarity
is govern by the same exponents in both cases, as revealed through the study of the time dependent particle
densities or the two-time autocorrelation. The same conclusion can be drawn when looking at the response to a 
perturbation that conserves the quantity $K$, see equation (\ref{eq:K}). 
It also remains true that the scaling properties of responses are found to depend on whether the perturbation conserves
the value of $K$ or not. Of course, other responses need to be studied in the future in order to further probe the universality of this
statement. 

All this points to a high degree of universality, suggesting that systems with reversible reactions form
excellent candidates for the experimental study of aging properties in reaction-diffusion systems. No fine-tuning
of the system parameters is needed in order to have slow dynamics, as demonstrated in experimental studies of
excited-state proton transfer reactions \cite{Hup92,Sol01,Pin01}. As our theoretical findings point to
a high robustness of the reported results, a future experimental verification of our predictions can be envisioned. 

Our present study can readily be generalized to other reaction-diffusion systems. In this paper we considered one of the
simplest reversible reaction scheme where two $A$ particles coalesce to form a $C$ particle which then can again decompose
into two $A$ particles. Other reaction schemes that can be studied are given by 
\begin{equation}
A + B  \rightleftarrows  C 
\end{equation}
or 
\begin{equation}
A + B \rightleftarrows C + D~.
\end{equation}
Especially the second reaction scheme is of interest, as it is not only realized readily in experiments
(a well known example is ethanoic acid dissolved in water that forms ethanoate and hydronium ions following
the reactions $CH_3CO_2H + H_2O \rightleftarrows CH_3CO_2^- + H_3O^+$), but it also has the interesting property
that {\it three} different quantities are conserved \cite{Elg08}. It would be interesting to study the different
responses in these systems in order to see how the scaling properties change when some or all conserved
quantities are broken by the perturbation. Work along these lines is planed for the future.

\ack
This work was supported by the US National
Science Foundation through DMR-0904999.

\end{document}